# Decoding Working-Memory Load During n-Back Task Performance from High Channel NIRS Data

**Christian Kothe**[1*], **Grant Hanada**[1], **Sean Mullen**[1*], **Tim Mullen**[1]
[1]Intheon, La Jolla, CA, United States
[*]**Correspondence:** christian.kothe@intheon.io, sean.mullen@intheon.io

## Abstract

Near-infrared spectroscopy (NIRS) can measure neural activity through blood oxygenation changes in the brain in a wearable form factor, enabling unique applications for research in and outside the lab. NIRS has proven capable of measuring cognitive states such as mental workload, often using machine learning (ML) based brain-computer interfaces (BCIs). To date, NIRS research has largely relied on probes with under ten to several hundred channels, although recently a new class of wearable NIRS devices with thousands of channels has emerged. This poses unique challenges for ML classification, as NIRS is typically limited by few training trials which results in severely under-determined estimation problems. So far, it is not well understood how such high-resolution data is best leveraged in practical BCIs and whether state-of-the-art (SotA) or better performance can be achieved. To address these questions, we propose an ML strategy to classify working-memory load that relies on spatio-temporal regularization and transfer learning from other subjects in a combination that has not been used in previous NIRS BCIs. The approach can be interpreted as an end-to-end generalized linear model and allows for a high degree of interpretability using channel-level or cortical imaging approaches. We show that using the proposed methodology, it is possible to achieve SotA decoding performance with high-resolution NIRS data. We also replicated several SotA approaches on our dataset of 43 participants wearing a 3198 dual-channel NIRS device while performing the n-Back task and show that these existing methods struggle in the high-channel regime and are largely outperformed by the proposed method. Our approach helps establish high-channel NIRS devices as a viable platform for SotA BCI and opens new applications using this class of headset while also enabling high-resolution model imaging and interpretation.

## Introduction

Improvements in fabrication and electronics integration have enabled compact and wearable near-infrared spectroscopy (NIRS) devices with 1000+ viable channels for use in neuroimaging (e.g., Zhao *et al*, 2021, Ban *et al*, 2021, and Anaya *et al*, 2023). These devices typically feature multiple source-detector separations and enable high-resolution brain imaging using methods such as high-density diffuse optical tomography (HD-DOT, e.g., Wheelock *et al*, 2018). They can also be used in freely moving and potentially ambulatory participants, opening new avenues for neuroscience research (e.g., Vidal-Rosas *et al*, 2021). However, it remains an open question to what extent the increased channel count of such devices enables higher performance brain-computer interface (BCI) applications, since the greater spatial resolution is traded off with higher data dimensionality, potentially lower signal to noise ratio per channel, and a potentially large number of bad channels, all of which pose unique challenges for machine learning. While several studies have investigated high-resolution NIRS for BCI purposes (e.g., Shin *et al*, 2017 and Ang *et al*, 2014), with somewhat mixed results, there exist few, if any, studies that investigate the efficacy of the current generation of high-resolution devices, which have on the order of 2-10 times the channel count of those previously studied (2000-4000 vs. 200-1000). Another question is to what extent the resulting decoding models exhibit improved localization, for example, more focal spatial features in their weight maps, and whether such properties may aid model interpretability.

Here we analyze a high-density dataset collected with the recently developed high-channel wearable continuous-wave NIRS system proposed by Anaya *et al* (2023) known as Spotlight (Meta Reality Labs, California), shown in figure 1, while participants performed the n-Back working memory task (Kirchner *et al*, 1958) along with another cognitive task not analyzed here. The n-Back task is the most frequently employed experimental task design to elicit working-memory load in both NIRS and functional magnetic resonance imaging (fMRI) literature. The task often serves as a proxy for general mental workload in NIRS human-computer interaction (HCI) and human-factors studies (e.g., Herff *et al*, 2014, Huang *et al*, 2021); although, in neuroscience, workload tends to be understood as a more differentiated construct that separates among others perceptual load and cognitive load (e.g., Lavie, 1995, 2010). Nevertheless, working-memory load remains one of the most frequently investigated components of cognitive load, besides executive functions such as task-switching load, inhibitory control, and others. Several of these processes activate overlapping cortical areas. These areas include the left and right dorsolateral prefrontal cortex (DLPFC) and parietal areas (see e.g., Niendam *et al* (2012) for a good overview using an activation likelihood estimate (ALE) meta-analysis of diverse cognitive tasks, and Owen *et al* (2005) for an n-Back specific treatment, also using ALE, both in fMRI). Similar brain areas, including the DLPFC have also been implicated in NIRS research (see e.g., Cui *et al* (2011) for a comparison between fMRI and NIRS, and Fishburn *et al* (2014), where the left and right DLPFC were the only cortical areas showing statistically significant changes in the NIRS signal during a working-memory experiment).

Here we present and study the behavior of a generalizable machine learning approach, aiming to decode the participant's working-memory load level that is specifically tailored to high-channel NIRS data. Our approach represents a passive BCI (Zander and Kothe, 2011) with potential applications in HCI, human factors and occupational safety research using this new generation of devices. We also provide a comparison with several methods reproduced from the literature on the same data, which shows that the tested methods do not reach the same performance and are likely not optimally leveraging the high-channel data. We further include analysis of block averaged task-related hemodynamic responses as well as cortical source localization using HD-DOT to put our findings in context and to support the interpretation of the machine learning (ML) results.

## Materials and Methods

### Participants

The dataset analyzed in this study comprises data from 43 participants (20 female, 23 male) recruited from the general population via classified ads. The human-subject research was conducted in accordance with the principles embodied in the Declaration of Helsinki and local statutory requirements and was approved under WIRB Protocol #20190255. All subjects gave informed written consent. Given that the goal was to evaluate performance on as general a population as possible, the recruitment criteria were participants ages 18-65 with no exclusions other than pregnancy, relevant health conditions (i.e., recurring migraines, claustrophobia, hypersensitivity or allergy to plastic on skin contact, on medication for seizures or beta blockers, etc.), or permanent hair pieces or dreadlocks. None of the 92 respondents were excluded due to hair type or any other criteria. Of the 45 who self-booked and eventually attended a session, two were unable to complete the experiment and their data was excluded, leaving a total of 43 included in the study.

The age grouping of participants was 12 of 18-29 years, 19 of 30-49 years, and 12 of 50+ years. Participants' hair opacity was split with nearly half (19) having dark-brown hair, with the remainder relatively evenly spread out along a spectrum from "none" to "black". In terms of hair thickness and length, participants were almost evenly divided between "thin", "medium" and "thick" (13, 16 and 14 participants, respectively) and between "short", "medium" and "long" (15, 15 and 11 participants, respectively). Three were left-handed, and all had normal or corrected-to-normal vision and reported no health conditions relevant to their task performance. All but two participants had no prior experience with the n-Back task.

We performed separate analyses of both the full dataset (which we label the "ALL" dataset in subsequent analyses) and a smaller subset of 36 sessions (the "OK" subset) meeting pre-determined thresholds for data quality and participant task proficiency. The criteria for the OK subset were at most 15% coefficient of variation (CoV) of raw intensity on average across all channels, a measure of signal quality, which was not met in five of 43 sessions, and a minimum participant proficiency of at least 85% accuracy at the n-Back task at the n=2 difficulty level, which excluded an additional two sessions.

## Experimental Setup

Data was collected with a high-density modular Spotlight CW-NIRS device in a two-module bilateral forehead configuration, as shown in figure 1. The modules were centered on the F3 and F4 locations on the international 10-20 system, which are located approximately over left and right DLPFC, respectively, although the modules overlap additional adjacent prefrontal areas, including parts of the ventrolateral prefrontal cortex (VLPFC). These module sites were chosen in agreement with prior working-memory load literature (e.g., Owen *et al*, 2005), here focusing only on the prefrontal areas.

Each module has a dense hexagonal grid of 41 sources and 39 dual-wavelength detectors (680 and 850 nm) with all-to-all channel connectivity within the module, yielding, per module, 3198 channels across the two wavelengths. See also Anaya *et al* (2023) for more details. NIRS sampling rate was 6.98 Hz. Data was collected in a darkened room with the lab streaming layer (LSL, Kothe *et al*, 2012) and NeuroPype (Intheon, La Jolla, CA), on a desktop computer running Linux.

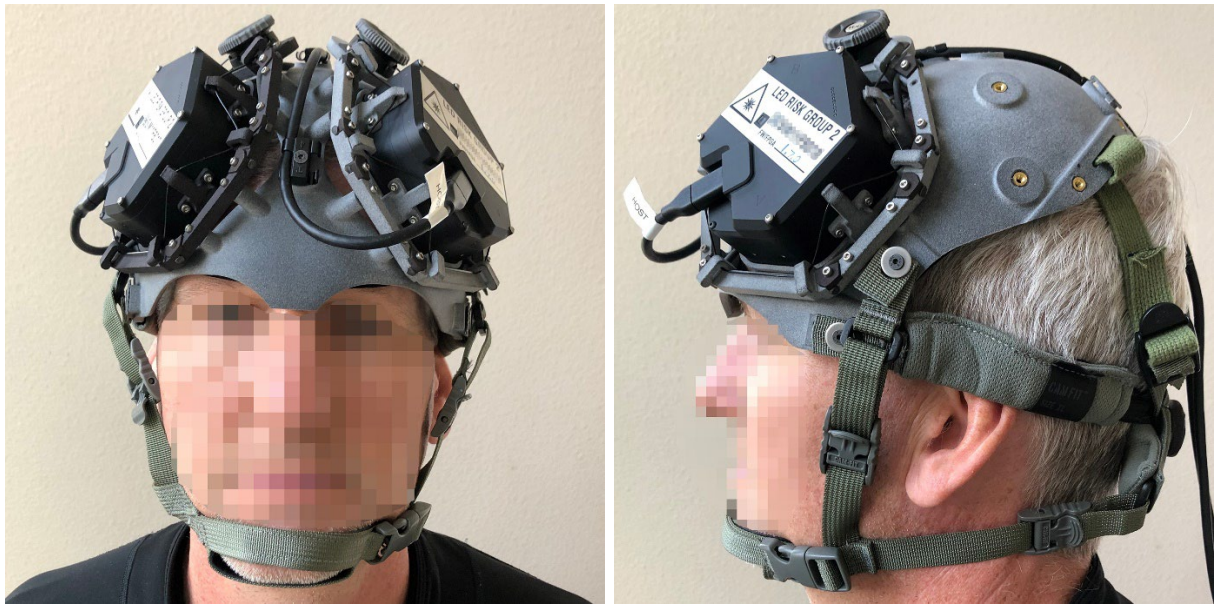

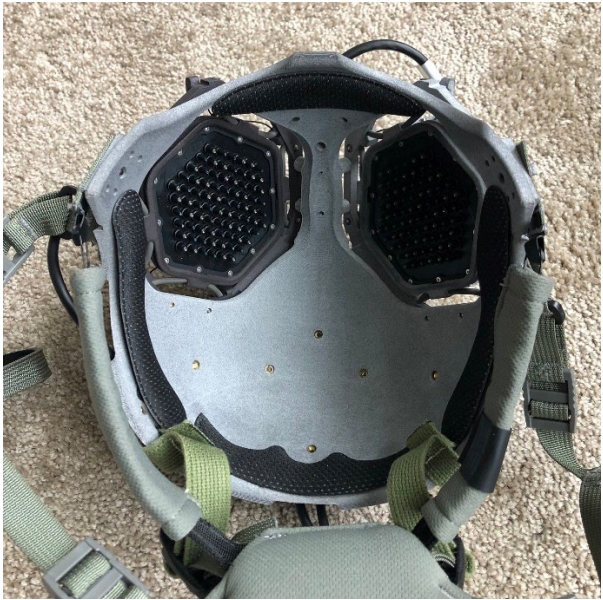
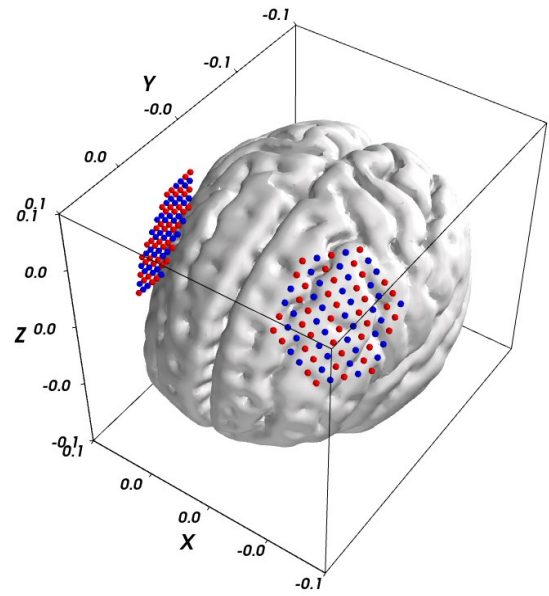

Figure 1. *Bilateral forehead configuration of the NIRS headset as used for the study. The helmet layout was adapted for the study and was 3d-printed in biocompatible Nylon PA12 (MJF) for the three head circumferences 54, 56, and 58 cm. Approximate optode locations (actual depth dependent on adjustable pressure setting) are shown in the bottom left with sources shaded in red and detectors in blue.*

## Experimental Task

The experiment session was partitioned into a series of 4 sets of tasks separated by self-paced breaks of at least 30 s as shown in figure 2. Each set consisted of three blocks of the n-Back task (3 minutes each, Kirchner *et al*, 1958) interleaved with two blocks of the Montreal Imaging Stress Task (MIST) task (2 minutes each, Dedovic *et al*, 2005), of which we only analyze the n-Back task in this article. Each n-Back block was further subdivided into three successive trials of 55 s duration each (all of same n), and each such trial consisted of 20 successive stimulus presentations (6 of them targets) using 18 consonants of the Latin alphabet, excluding X and Z as stimuli (X was reserved for the 0-back condition as the target letter). Letters were presented in black on a dim grey background in the center of the screen. We used an inter-stimulus interval of 1.5 s and stimulus duration of 0.5 s, totaling 40 s task performance per trial, preceded by a 10 s baseline and 5 s task instruction. For each stimulus presentation, participants were asked to respond to the stimulus only if that stimulus matched the one presented n items prior (e.g. 1-back match is if current letter is same as the previous letter shown, 2-back match is if current letter is same as the letter shown 2 letters back, while the special case of 0-back match is if the current letter is X), using a right-hand down-arrow keystroke. Conditions for n were limited to 0, 1, 2 and were balanced and pseudo-randomized across blocks, sets, and participants, yielding 36 total trials per session. The session was preceded by a brief practice period during which participants could familiarize themselves with the task's instructions. The task was presented using the Experiment Recorder software (Intheon, La Jolla, CA).

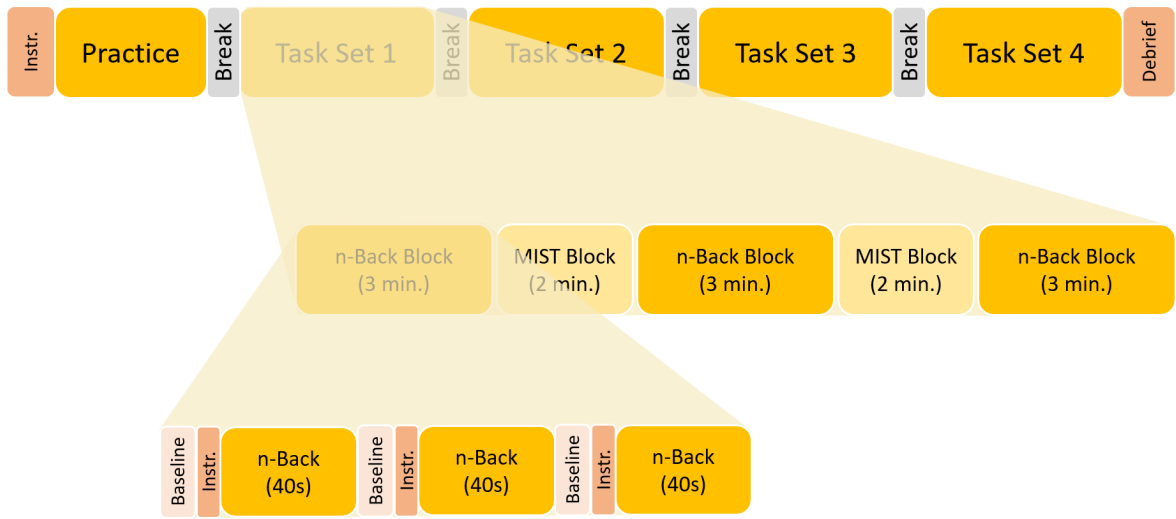

Figure 2. *Overview of experiment structure. Top row shows the timeline of the entire session from instruction to debriefing. Middle row shows the makeup of a single task set (same for each task set but with pseudo-randomized difficulty levels, i.e., n in n-Back), and bottom row shows the timeline of a single n-Back block within a given task set, which consists of 3 task performance periods (NIRS trials) each preceded by instruction and baseline, all same n. Task performance blocks themselves each comprise a succession of 20 letter stimulus presentations.*

## Decoding Pipeline

In the following we discuss the machine learning decoder employed to classify the n-Back workload level at single-trial granularity (where a trial is the 40 s task performance period, plus the preceding 15 s rest and instruction period). The decoding pipeline, whose steps and their order are summarized in figure 3 (bottom), proceeds as described in the following section. The training pipeline (top of figure) is described thereafter. These pipelines include several steps not typically employed in NIRS to address the high dimensionality of the data. We will also discuss generalizations beyond the working-memory paradigm studied here.

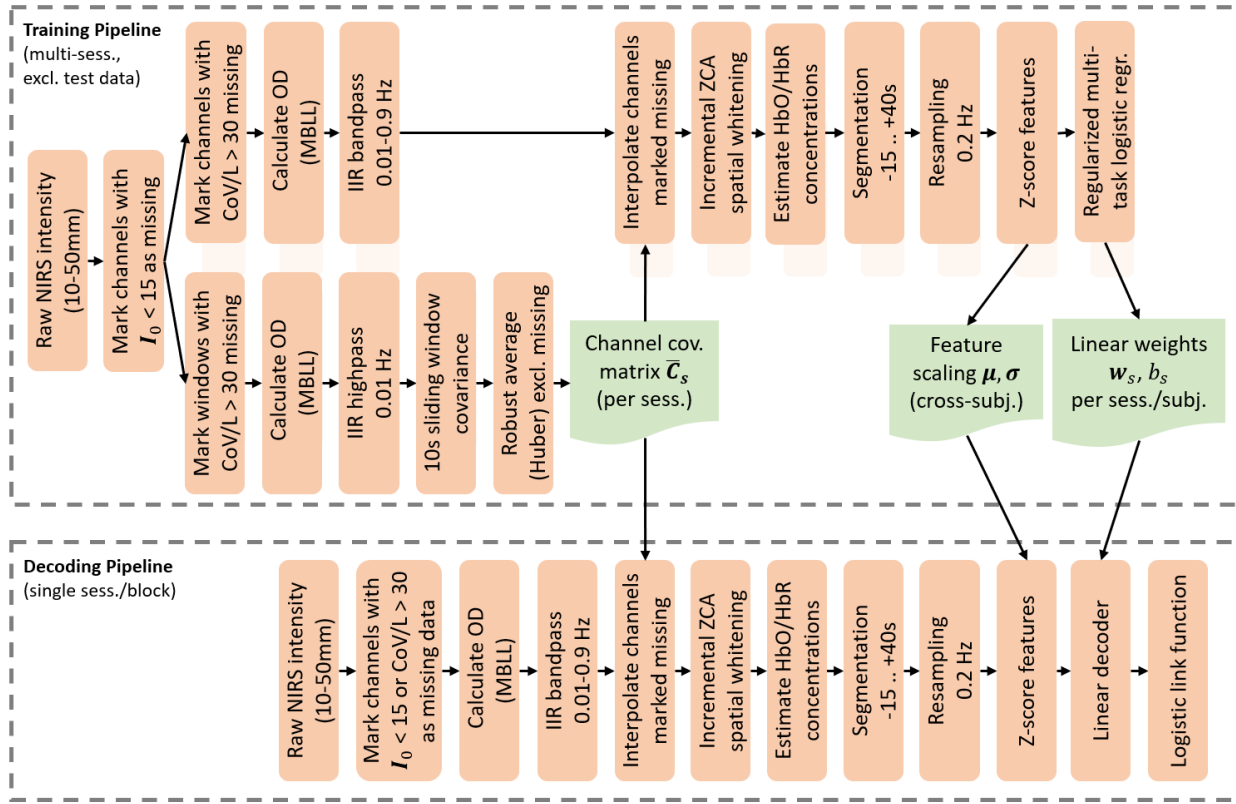

Figure 3. *Overview of the training (top) and decoding (bottom) pipeline. The three cards in the center with arrows into the decoding pipeline below indicate learnable parameters of the decoding model and remaining shaded boxes indicate processing steps. Abbreviations used: CoV/L is the coefficient of variation of a channel divided by its length (in cm), $I_0$ is the whole-session intensity average, OD is optical density, MBLL is the modified Beer-Lambert law, ZCA is zero-phase component analysis, and HbO/HbR are the oxy/deoxyhemoglobin chromophore concentration changes.*

*Raw intensity measures.* Our decoding pipeline operates on raw NIRS intensity, and begins with ambient light correction, that is, subtracting the per-detector ambient light readings provided by the NIRS device. We found that in the Spotlight device a combination of low-discrepancy ambient light readings, the presence of long channels, and brief channel illumination cycles result in some of the long channels reading out zero intensities after ambient light correction for a significant fraction of the session. To avoid generating infinities or spuriously large outliers in subsequent optical density calculation and z-scoring steps it was beneficial to shift the noise floor up by one quantization step by adding an offset of 1 (the analog/digital conversion unit) to the observed light levels, plus additive Gaussian noise with a standard deviation of ½. The latter has negligible impact on the signal to noise ratio (SNR) since the signal is integrated over longer (multi-second) time scales in our subsequent analysis. To reduce the input dimensionality, we further retain only channels of 10-50 mm source-detector separation, yielding a total of $n = 5540$ channels across the two wavelengths and both modules.

*Channel preprocessing and imputation.* Next, we remove (or mark as missing) all channels that have a sliding-window (5 s) coefficient of variation (CoV) > 30 in more than half of the session after normalizing (dividing) the CoV by the channel length (in cm). Similarly, we mark all channels for which the within-session average intensity $I_0$ is below 15 as missing.

Following that, we calculate the change in optical density (ΔOD) based on the modified Beer-Lambert law (MBLL, Delpy *et al*, 1988), and reference ΔOD to be relative to the within-session OD average. We then suppress low-frequency drifts and high-frequency physiological artifacts using a 5th order Butterworth infinite impulse response (IIR) band-pass filter with a passband of 0.01-0.9 Hz, using zero-phase (forward-backward) filtering.

Since our decoding models are trained on aggregate multi-session data, we impute, separately for each session $s \in \{1, \ldots, S\}$, the subset of channels previously marked as missing ("bad") for that session, whose indices we denote by the set $\mathrm{B} \subseteq \{1, \ldots, n\}$ (dropping subscripts $s$ to keep notation light) using a least-squares interpolation (e.g., Enders *et al*, 2022) as $\hat{\boldsymbol{o}}_\mathrm{B} = \boldsymbol{R}\boldsymbol{o}_\mathrm{G} + \boldsymbol{\varepsilon}$. Here we denote by $\boldsymbol{o}_\mathrm{B} \in \mathbb{R}^{|\mathrm{B}|}$ the vector of per-channel ΔOD measurements at a given time point restricted to the set of *bad* channel indices (and likewise for the set of *good* channel indices in $\boldsymbol{o}_\mathrm{G}$), and where $\varepsilon_i \sim \mathcal{N}(0, 10^{-6}\sigma_i^2)$ is additive Gaussian noise scaled by the respective channel's (post-interpolation) standard deviation $\sigma_i$ to keep the data full rank. We estimate the session's imputation/reconstruction matrix $\boldsymbol{R} \in \mathbb{R}^{|\mathrm{B}| \times |\mathrm{G}|}$ from a multi-session grand-average spatial covariance matrix $\overline{\boldsymbol{C}} \in \mathbb{R}^{n \times n}$ according to

$$\min_{\boldsymbol{R}} \left\| \overline{\boldsymbol{C}}^{1/2}_{:,\mathrm{B}} - \boldsymbol{R}\overline{\boldsymbol{C}}^{1/2}_{\mathrm{G},:} \right\|_2^2$$

where the notation $\overline{\boldsymbol{C}}_{:,\mathrm{B}} \in \mathbb{R}^{n \times |\mathrm{B}|}$ represents matrix $\overline{\boldsymbol{C}}$ with rows reduced to index set $\mathrm{B}$ (and analogously with $\overline{\boldsymbol{C}}_{\mathrm{G},:} \in \mathbb{R}^{|\mathrm{G}| \times n}$ for columns reduced to index set $\mathrm{G}$); this can be solved as a standard regression problem. $\overline{\boldsymbol{C}}$ is one of the trainable parameters of our method, whose estimation we describe in the Training Pipeline section. Since our data stems from two non-interacting probe modules, one per hemisphere, we process each module's channels independently to reduce computational overhead and improve the conditioning of the interpolation problem (this is equivalent to treating $\overline{\boldsymbol{C}}$ as block diagonal).

*Standardization.* A second key step in our multi-session preprocessing pipeline is to minimize covariate shift across sessions or participants, which we implement by spatially whitening each session, resulting in data of uniform covariance, as a simple type of domain adaptation (e.g., Pal and Sudeep, 2016), a domain being a session. To allow this standardization step to be potentially applied in real time using causal processing (although in this analysis the IIR filter and ΔOD re-referencing were non-causal) all results were generated using an incrementally updated zero-phase component analysis (ZCA, Bell & Sejnowski, 1997). This method recursively updates a time-varying within-session covariance matrix $C_t \in \mathbb{R}^{n \times n}$ at time $t$ using the formula

$$C_t = \frac{1}{t}((t-1)C_{t-1} + \hat{o}_t \hat{o}_t^\top)$$

Where $\hat{o}_t \in \mathbb{R}^n$ is the ΔOD sample at time $t$ after imputation has been applied as described earlier. The (spatially) zero-phase whitening transform $W_t$ is updated as $W_t = C_t^{-1/2}$, using the matrix square root, after each trial only (to minimize within-trial variation), and each sample $\hat{o}_t$ is whitened as $\tilde{o}_t = W_t \hat{o}_t$. Due to the high channel count we found it necessary to use ZCA with shrinkage regularization towards a diagonal covariance matrix using the formula $\tilde{C}_t = (1-\lambda)C_t + \lambda \mathrm{diag}\, C_t$, here using $\lambda = 0.5$, although in prior work we had found values as low as $10^{-6}$ to be feasible. Lastly, to fully leverage available training data in our within-session blockwise cross-validation, we first initialize the ZCA filter on the respective training blocks of a given session and then continue incremental processing on the held-out test block.

*Feature Extraction.* Next, we extract linear signal features by first estimating a time series of concentration changes for oxyhemoglobin (ΔHbO) and deoxyhemoglobin (ΔHbR) chromophores, assuming a differential pathlength factor of 6, as $\tilde{c}_t = A^+ \tilde{o}_t$ where $A^+ \in \mathbb{R}^{n \times n}$ is the Moore-Penrose pseudoinverse of a block-diagonal sensitivity matrix of appropriately rescaled wavelength- and chromophore-specific extinction coefficients. However, we caution that the individual wavelength signals have at this stage been mutually decorrelated by the ZCA filter, so the resulting estimates $\tilde{c}_t$ are not strictly identifiable as HbO and HbR[1]. We then extract time segments around each n-Back trial starting 15 s prior to task performance onset and ending 40 s after task performance onset. This leaves no gap between successive trials, but at evaluation time we ensure adequate train/test set separation by using a block-wise cross-validation that splits data at experiment breaks (described in the Evaluation Scheme section).

We then down-sample each extracted trial segment to 0.2 Hz using polyphase resampling, which yields $T = 12$ temporal features per channel, and concatenate features across the $W = 2$ wavelengths (or chromophores) and each of the $C = n/2$ dual-wavelength channels, resulting in a resampled feature vector $r_i \in \mathbb{R}^D$ in a $D = 66{,}480$ dimensional feature space for each trial $i$. Due to the relatively high temporal resolution of these features, a linear classifier can learn to extract information equivalent to a broad class of commonly used handcrafted temporal features such as time-window averages, slopes, wavelets, or pre-stimulus baselines from these data, and is therefore relatively task independent without the need for custom feature engineering but is unable to replicate some non-linear features such as ratios or moment statistics. Since our data comprises 1000s of channels measuring highly overlapping brain areas that are linearly related, we assume here that the latent chromophore concentration changes in the brain volume are likely best approximated by linear combinations of multiple channels. These cannot be recovered after non-linear operations have been applied to the per-channel concentrations, and therefore we restrict ourselves here to linear feature extraction only. The feature vector for each trial is then z-scored using training-data statistics $\mu, \sigma \in \mathbb{R}^D$, yielding z-scored features $x_i = (r_i - \mu)/\sigma$.

---

[1] Applying ZCA *after* concentration appeared to perform less well in this and a forthcoming study.

*Linear Decoding.* The resulting data are then mapped through a linear decoder stage as $\hat{y}_i = w_s^\top x_i + b_s$, which we train using a type of heavily regularized logistic regression (discussed in the Training Pipeline section), followed by a logistic link function $p_t = 1/(1 + e^{-\hat{y}_i})$ to estimate the probability that the trial is of one class vs the other.

*Real-Time Usage.* While this pipeline was not employed in real time in this study, all steps are easily configurable for low-latency real-time operation, using, for example, a causal elliptical IIR filter, or the moving average convergence divergence filter as in Cui *et al* (2010) in place of the Butterworth IIR filter. We have tested both with our implementation.

## Training Pipeline

In the following we describe the training scheme used to calibrate the decoder, which is depicted in the diagram in figure 3 (top row). We estimate all trainable parameters of the pipeline based on a data corpus that, in the general case, comprises data from multiple participants and/or sessions, including from the target subject, and which may include *training* portions of the target session (to enable within-session cross-validation), but which strictly excludes any data used for *testing*. The approach simplifies when no data from the target subject or session is included (results for both approaches are presented).

*Channel Covariance Matrix.* Due to the need for aligned multi-session data, the first training step is to estimate a bad-channel imputation model for the montage at hand (see Decoding Pipeline), which in our approach relies on the grand-average spatial covariance matrix $\bar{C}$. This matrix is estimated as a robust Huber mean (Huber, 1996), which we take over a collection of successive short-window (10 s) covariance matrices $C_w$ pooled across all training sessions, using Huber parameter $\delta = \eta \hat{\sigma}$ where $\hat{\sigma} = k \, \text{median} \, \|C_w - \text{median} \, C_w\|_F$ is a robust estimate of the dispersion among the covariance matrices ($k \approx 1.4826$ is a scale factor that appears when approximating a standard deviation from a median absolute deviation, and $\|\cdot\|_F$ is the Frobenius norm of a matrix) and $\eta$ is an outlyingness threshold in standard deviations, which we fix at 3/4.

To further minimize the impact of artifactual training data portions (channels, time periods) entering this robust average, we mask, in 5-second successive windows, any window of any channel where the CoV exceeds 30 (again after correction for channel length) as missing data, or the entire channel within a given session if its average intensity is below 15, and use masked median and Huber mean estimators to obtain $\overline{C}$. Note that these masking criteria and their parameters have equivalents in the bad channel rejection step in the decoder pipeline discussed earlier, and the covariance matrices are estimated on otherwise identically preprocessed per-session optical density data, except for using a 0.01 Hz highpass filter (5th order Butterworth IIR) rather than a bandpass filter prior to covariance estimation. The latter was chosen to allow the estimator to capture the channel correlation structure across a broader frequency range, and to increase the effective degrees of freedom in the average.

To strictly avoid any statistical "double dipping" (c.f., Kriegeskorte *et al*, 2009), we estimated matrices $\overline{C}_s$ used for imputing channels in each target session $s$ from all *other* sessions, excluding that target session, for all results reported herein. A computationally less expensive alternative would be to empirically estimate a single matrix on data from a prior study or high-quality reference dataset that uses the same montage, when available, although we found that simply using $\overline{C}$ does not appear to inflate performance.

*Linear Classifier.* To train the linear classifier, we first preprocess the data in each training session identically to the scheme used at test time to ensure optimal train/test match (cf. figure 3 top row), and perform feature z-scoring using per-feature statistics $\boldsymbol{\mu}, \boldsymbol{\sigma}$ taken across all pooled training(-only) trials across participants/sessions. The classifier is then trained with a multi-task logistic regression objective (e.g., Zhang and Yang, 2021) where we treat the data from each session as a separate task, and where we use a centered multi-task learning (MTL) formulation (based on McDonald *et al*, 2016) to encourage models for each subject to be similar to a common latent model (such that the deviation from this model has a low $l_2$ norm). We combine this with a multi-task feature learning (MTFL) cost function (as in Argyriou *et al*, 2006) to learn a low-rank subspace of spatio-temporal features that encourages deviations from the common model to lie in a shared subspace across sessions. Lastly, we further reduce the effective degrees of freedom of the model using a spatio-temporal smoothness regularization that we implement with additional Tikhonov regularization terms (Tikhonov *et al*, 1995). In the following we denote by $\boldsymbol{X}_{si} \in \mathbb{R}^{C \times WT}$ the z-scored feature vector $x_{si}$ for subject $s$ and trial $i \in \{1, \ldots, N\}$ rearranged into a matrix, along with its associated class label $y_{si} \in \{0,1\}$. Solving the (jointly convex) cost function

$$\min_{W_s, b_s} \frac{1}{SN} \sum_{s=1}^{S} \sum_{i=1}^{N} \log\left(1 + e^{-y_{si}(\langle W_s + W_0, X_{si} \rangle_F + b_s)}\right) + \alpha \|[W_1, W_2, \ldots, W_S]\|_*$$

$$+ \sum_{s=0}^{S} \beta \|\Gamma_U \text{ vec } W_s\|_2^2 + \gamma \|\Gamma_V \text{ vec } W_s\|_2^2 + \|W_s\|_2^2$$

estimates the weight matrices $W_s \in \mathbb{R}^{C \times WT}$ and intercept $b_s$ for each session $s$, along with the weights of a shared cross-session model $W_0$ as in centered MTL. The matrices $\Gamma_U$ and $\Gamma_V$ are sparse spatial and temporal Tikhonov operators, respectively. The form $\|\cdot\|_*$ denotes the trace norm used to realize the low-rank subspace (MTFL) assumption and $\langle \cdot, \cdot \rangle_F$ is the Frobenius inner product. The parameters $\alpha$, $\beta$, and $\gamma$ are the cross-session coupling strength and spatio-temporal regularization parameters, respectively. If no data from the target subject is included, the problem simplifies, and all combinations of $W_*$ matrices each reduce to a single occurrence of $W_0$ and $b_s$ becomes $b_0$.

The temporal Tikhonov operator is formed as $\Gamma_V = T \otimes I_n$ where $\otimes$ is the Kronecker product, $I_n \in \mathbb{R}^{n \times n}$ is an identity matrix, and where $T \in \mathbb{R}^{T \times T}$ is a banded matrix with elements:

$$T = \begin{pmatrix} 2 & -1 & & 0 \\ -1 & \ddots & \ddots & \\ & \ddots & \ddots & -1 \\ 0 & & -1 & 2 \end{pmatrix}$$

The spatial Tikhonov operator is likewise formed as $\Gamma_U = I_{WT} \otimes L$ where $L \in \mathbb{R}^{C \times C}$ is a normalized spatial Laplacian operator that is derived as

$$L = \frac{\Omega - D}{1 - \sum_{ij} \Omega_{ij}/C}$$

where $D \in \mathbb{R}^{C \times C}$ with $D_{ii} = \sum_j \Omega_{ij}$ is a diagonal matrix, $\Omega \in \mathbb{R}^{C \times C}$ is a positive definite spatial taper matrix with elements $\Omega_{ij} = \phi_{3,1}(d_{ij}/r)$ where $d_{ij}$ is the Euclidean distance between the spatial midpoints $m_i$ and $m_j$ of channels $i$ and $j$; $r$ is a smoothing radius (here set to 15 mm), and $\phi_{3,1}(x)$ is a radial cutoff polynomial in 3-dimensional space as defined by Wu *et al* (1995) that results in $L$ being sparse. To account for the skull curvature, we displace the linear channel midpoints for channel $c$ perpendicular to the scalp using a surface normal $n_c$ at the respective midpoint using the formula

$$m_c = \frac{s_c + d_c}{2} - n_c \frac{\|s_c - d_c\|}{\tau}$$

where $s_c$ and $d_c$ and are the source and detector positions for channel $c$ and $\tau = 3/2$ is a proportionality factor that governs the ratio of lateral to radial (depthwise) smoothness of the regularizer. Depending on the montage geometry, the surface normal can be obtained from a coregistered scalp mesh, point towards the head center, or be chosen as perpendicular to the local optode plane, which was used here. Once $W_s$ and $W_0$ have been estimated we can then derive the linear decoder weight vector for a given session as $w_s = \text{vec}(W_s + W_0)$.

We solve this optimization problem using first-order accelerated proximal gradient decent (e.g., Nesterov 2007, Beck & Teboulle, 2009), which converges in ca. 5 minutes on an RTX 3080-class GPU using CuPy (Okuta *et al*, 2017) for a given set of parameters $\alpha, \beta,$ and $\gamma$. The hyper-parameters (spatial smoothness β, temporal smoothness γ, and low-rank coupling parameter α) were optimized in a 3-fold blockwise nested cross-validation on the respective training set using the test-based population size adaptation (TPBSA) method implemented in Nevergrad (Oquab *et al*, 2019), and no manual tuning was performed (see also Evaluation Scheme section for more detail on the cross-validation). When learning parameters for a new session $s + 1$, the parameters for the preceding sessions, if the prior session pool is large enough, may also be held fixed to reduce the size of the optimization problem.

To recap, the main task-specific parameters of this pipeline are the segment length relative to a stimulus (here -15 to 40 s), the bandpass filter (0.01-0.9 Hz) and the temporal resolution (here 0.2 Hz), and to a lesser extent the artifact removal settings, while the montage-specific parameters are the ZCA regularization parameter $\lambda$ (0.5), the smoothing radius $r$ (15mm) and the anisotropy factor $\tau$ (3/2), while all other parameters are learned from the data. The logistic loss may be replaced by the square loss (also dropping the logistic link) to utilize the method in a regression context.

## Comparison with Existing Methods

To benchmark decoding performance on our dataset, we reproduced three methods from prior literature which we selected due to their use of high-channel data (Shin *et al*, 2017 and Ang *et al*, 2014), or high reported performance (Kesedžić *et al*, 2021). We tested these methods using an evaluation scheme analogous to our proposed method (described in Evaluation Scheme), but several algorithmic adaptations were necessary to match these methods to our NIRS montage (which exceeds the channel count of most prior use cases) and our task timing, as described in the following.

The method of Shin *et al* (2017) was reproduced as closely as possible given the differences in experimental task timings and montages. We retained channels of 10-20 mm and 25-40 mm source-detector separation and utilized a minimum intensity threshold of 15. We used two separate time windows, one covering the pre-stimulus baseline at -5 to 0 s, and another covering the task performance period at 5-40 s post-stimulus for temporal features. All other details of the method were matched exactly (CoV-based channel rejection criterion, bandpass filter, and hierarchical shrinkage linear discriminant analysis (sLDA) classifier).

Since Ang *et al* (2014) use similar task timings as our study, we were able to match their method almost exactly, including the channel selection criterion, IIR lowpass filter, detrending, ΔOD offset, common average reference, differential features, support vector machine classifier, and mutual information based feature selection. The only difference is that we retained 4 times as many features (k=40) than the authors had found optimal since our montage has ca. 4 times as many (but potentially lower-SNR) channels of the selected range (using their k=10 retained features performed less well in experiments).

We reproduced the method by Kesedžić *et al* (2020) to the extent possible given the differences in montage density and experimental task timings. The method is technically channel density independent since it relies on multi-channel averages. However, the artifact removal operates on a single-channel basis, and was found to hamper performance when enabled, possibly due to frequent changes of the channel inclusion mask over time and the resulting signal discontinuities; we therefore report here the (better) results without that artifact removal step. For the left and right DLPFC channel averages in Kesedžić's method, we selected channels whose linear midpoint falls within a 20 mm radius around the F3 and F4 locations in the 10-20 system. The time window for feature extraction was set to 0-40 s instead of 0-75 s to adapt the method to the duration of our shorter tasks. We found that sLDA with cross-validated shrinkage parameter (e.g. Peck and van Ness, 1982) performed better than the sparsifying shrunken centroids regularized discriminant analysis method (SCRDA, Guo *et al* 2007) used by the authors, possibly since the latter method's extra sparsity is made somewhat redundant by the sequential feature selection that is also used, and we report these better results here.

Common to all tested methods, we add one integer unit to the raw intensity to avoid very low intensities in the optical density calculation, which in most cases improved performance somewhat.

## Evaluation Scheme

We compare two types of methods here, subject-*independent* methods (Kesedžić *et al*, and a subject-independent variant of our proposed method), which require no training data for the target subject, and subject-*specific* methods, which are adapted using some amount of training data for the target subject, and potentially using other participants' data as well (Shin *et al*, Ang *et al*, and our proposed method). The two subject-independent methods were evaluated using a rigorous cross-validation (CV) that proceeded in a leave-one-subject-out fashion, using no data from the target subject. The subject-specific methods were evaluated using a 4-fold blockwise cross-validation on each subject's session, whose test partitions correspond to the 4 experimental task sets (separated by experiment breaks), since this splitting scheme is considered less likely to overestimate performance than a randomized CV on neural data (e.g., Varoquaux *et al*, 2017).

For our proposed method in its subject-*specific* variant, which uses multi-task learning, the within-subject cross-validations were computed side by side in parallel, but otherwise using the same data splits and full train/test separation within each subject/session (and using a different order of the 4 task sets for each session). Hyper-parameters were optimized on the respective training sets only in a rigorous nested cross-validation (e.g., 3-fold to match natural experiment breaks between the 3 training task sets).

## Block Averaging

To visualize average hemodynamic responses in the three different n-Back conditions, we use a block averaging approach of source-resolved NIRS data. To mitigate artifacts, we first subtracted ambient light measures, shifted the noise floor by one integer quantization unit, retained channels in the 10-50 mm source detector separation range, estimated delta-optical density using the modified Beer-Lambert law, then applied temporal derivative distribution repair (TDDR, Fishburn *et al*, 2014) for motion artifact reduction, and lastly applied 5th order Butterworth IIR high- and lowpass filters with band edges at 0.05 and 0.2 Hz, respectively. We found that when using the default TDDR method, larger evoked activity can be significantly suppressed, likely since such activity appears to violate the method's assumption that clean data are approximately Gaussian distributed. However, we accepted this tradeoff here to ensure that potential outliers are sufficiently suppressed in the block averages.

The resulting artifact-reduced data were then imaged using an HD-DOT workflow (e.g., Tremblay *et al*, 2018) and HbO/HbR concentration changes were estimated in the same step. Specifically, we coregistered each of the three cap sizes to the MNI152 (Fonov *et al*, 2009) volume and cortex meshes using affine transformation, computed forward model Jacobian matrices based on a 4-layer finite element model using the Toast++ software (Schweiger and Arridge, 2014), and calculated inverse estimates for HbO/HbR source concentration changes using the Rytov approximation (e.g., Madsen, 2012) using a multi-spectral basis function set.

Lastly, we segmented the NIRS data locked to task performance onset with a 5 s pre-task baseline subtracted, and then averaged all blocks for each of the three n-Back conditions. All computation other than the Jacobian matrix computations as mentioned above was done using NeuroPype on a desktop computer running Linux.

# Results

## Model Performance

We compared the performance of the three previously described subject-specific methods, and two subject-independent methods, and evaluated their accuracy on a binary classification task between any two n-Back workload levels. We tested all three possible subsets of our experimental data (n=0 vs n=1, n=0 vs n=2, and n=1 vs n=2), and the analysis was repeated on the 43-sessions ALL subset (no data exclusions at all besides the two incomplete sessions), and on the 36-session OK subset. Results are shown in figure 4.

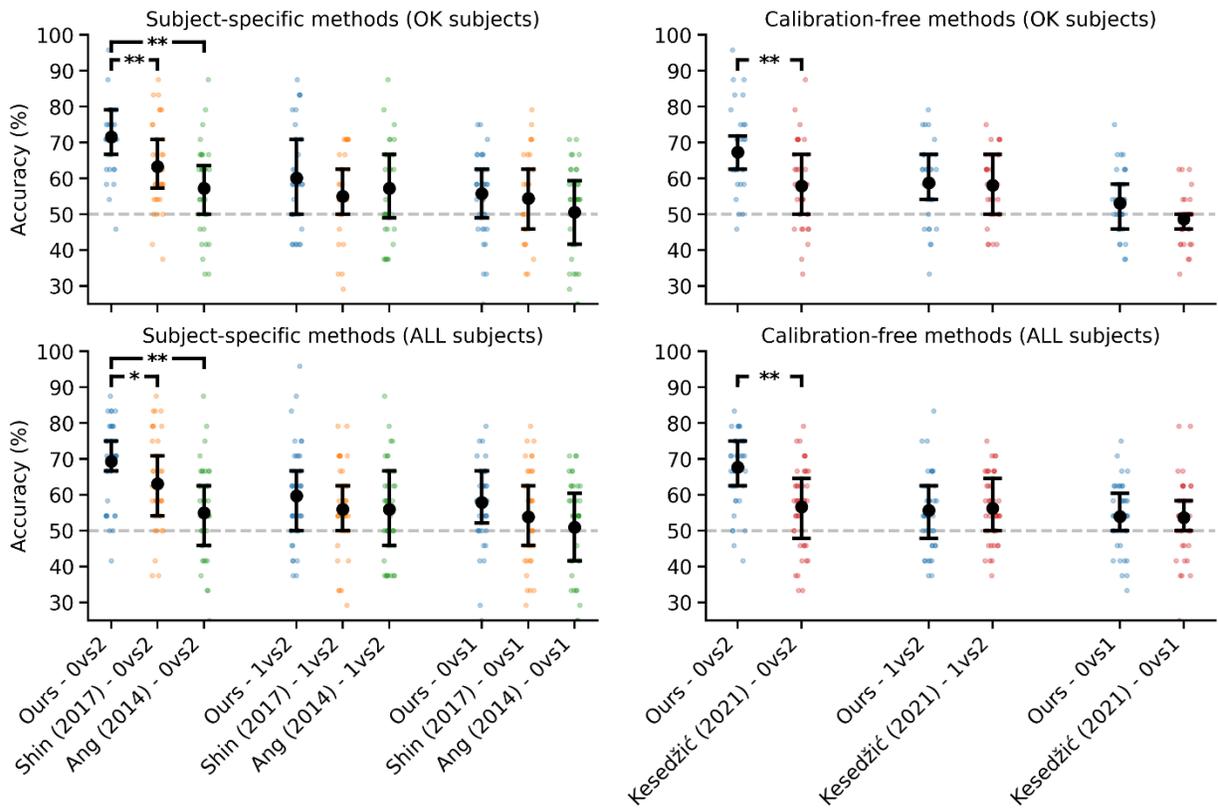

Figure 4. *Performance of tested methods, broken out by subject subset (OK participants in top row, ALL participants in bottom row), and methods grouped into subject-specific (left column) and subject-independent ("calibration-free", right column) methods. Average performance is indicated by a black dot, vertical brackets represent the interquartile range. Performance on individual sessions (36 for the OK subset and 43 for the ALL subset) is overlaid, colored by*

*method. Chance level is indicated by a dashed line at 50% accuracy. Stars indicate the significance level of (Bonferroni-corrected) paired t-tests between the proposed method and the respective other method(s) in each group.*

The figure shows that the proposed method achieves significantly higher accuracy than the other tested methods at decoding the n-Back workload level for the two-step n=0 vs 2 difference, using paired t-tests with Bonferroni multiple comparison correction accounting for the number of subject subsets (2) and task conditions tested (3). Similar statistical significance levels were also obtained when using Wilcoxon signed rank tests. This improvement over other tested methods holds regardless of whether the method was used in a subject-specific or subject-independent configuration, and regardless of whether only good sessions were retained or whether all sessions were included in the analysis. Best performance was obtained in the subject-specific configuration, where our method yielded 71.5 +/- 9.2% accuracy on the OK session subset and 69.3 +/- 10.7% on the ALL subset. However, performance of the subject-independent method is not much worse with 67.4 +/- 10.9% on the OK session subset, and 67.7 +/- 10.4% on the ALL subset. The slight increase when using the ALL subset may be due to the larger training set size and indicates that the method does not appear to suffer appreciably when artifactual or otherwise low-quality data are included in the training data.

For the more difficult to decode one-step load level differences (n=1 vs 2 or n=0 vs 1), our method tends to produce higher decoding accuracies than other methods in all tested setups except when comparing to Kesedžić on all participants for n=1 vs 2 (figure 4 bottom right panel, middle test), although performance differences in these one-step load level contrasts did generally not survive multiple-comparison correction due to the large performance range across participants. We can see that the n=0 vs 1 load level difference appears to be harder to decode accurately for all methods than n=1 vs 2, possibly since both n=0 and n=1 do not engage working memory much, and performance of all methods is below 60%. Also of note, the performance gap between the proposed method and the next-best method is not constant but appears to scale roughly proportionally with the separability of the data (i.e., with a larger n difference, a cleaner session subset, and on data where the other methods themselves fare relatively better).

## Spatio-Temporal Weights

In the following we review the weight tensor for the model with highest performance (n=0 vs 2, trained on the OK subject subset). Due to the large number of weights across a wide range of channel lengths, time points, and high spatial resolution, we focus here mostly on temporal and spatial slices through the model, integrating the respective other dimensions out, the results of which are shown in figure 5 and figure 6.

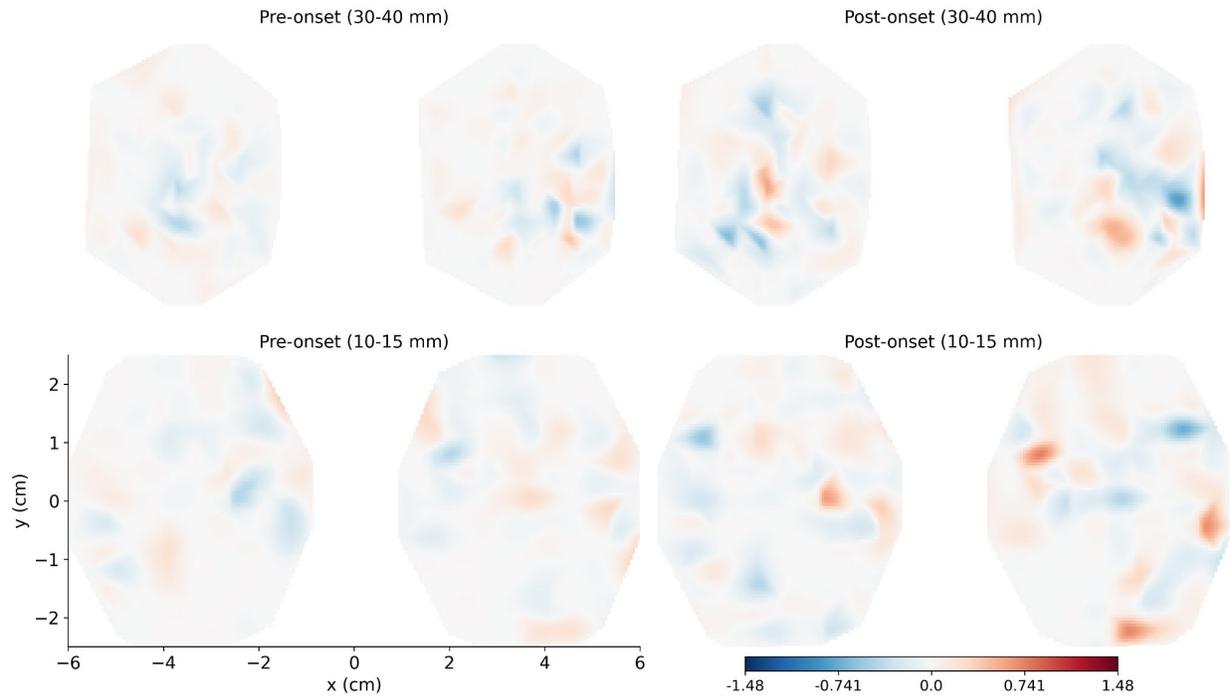

Figure 5. *Model weights projected on each of the left and right hemisphere modules for the 0 vs 2-back model summed across different subsets of channel source-detector separations (30-40 mm channels top, 10-15 mm short channels bottom) and time slices (pre task onset on the left, post task onset on the right). The 2D map is shown as if the view is facing the top of the subject's head, face looking in the downwards direction (left hemisphere is on the right), with montage geometry unwrapped into a plane. Full model represents 10-50 mm channels and -15 s to +40 s time (relative to task onset), and HbO/HbR. Note polarity inversions between pre vs post onset, and post-onset weight concentrations for 30-40 mm near module center (over DLPFC) and towards the edge on the left hemisphere (DLPFC/VLPFC), cf. also* figure 1 *for spatial coregistration.*

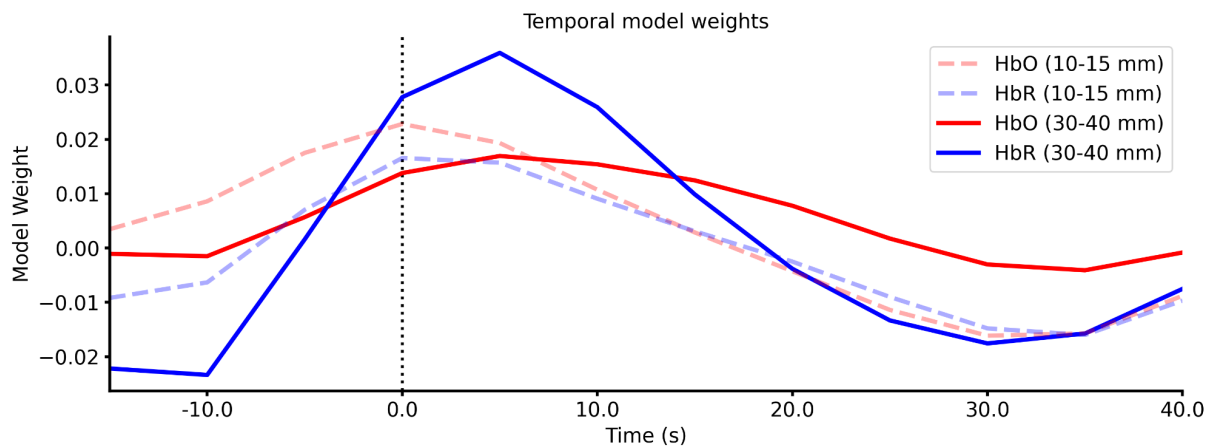

Figure 6. *Total model weight for the 0-2-back model shown across time and summed over different channel subsets. HbO/HbR are shown in red/blue, with short channel weight depicted by dashed lines and long (deep) channel weight as solid lines. Model is the same as in figure 5. Note strong negative weight on HbR prior to task onset followed by large peak immediately after task onset vs. a shallow positive peak for HbO that covers approx. the first half of the task duration then falls off towards the end of the task block.*

The weight figures generally show a larger post-onset weight compared to pre-onset for the same channels (left vs right column in figure 5) and some degree of polarity inversion between pre and post-onset maps that is compatible with the classifier having designed differential features (i.e., subtraction of a pre-task baseline from a during-task feature), although this is more clearly visible in the temporal view (figure 6). The model uses both long and short channels (top and bottom rows in figure 5, respectively) as one would expect if the model performed a form of short-channel regression, and exhibits a more diffuse weight pattern for short channels, and more focal pattern for long channels that is approximately central under the right-hemisphere module (F4 scalp location in 10-20 system) and central as well as laterally displaced on the left-hemisphere module (F3 and F5/FC5 scalp locations). These weight hotspots are approximately above the DLPFC and left VLPFC, which is in line with expectations; see also figure 1 for coregistration with cortex.

The temporal weights (figure 6) show a several-second peak shortly after task onset in HbR, and a much shallower HbO peak roughly covering the first two thirds of the task period, as well as a negative (differential) weight on the pre-task baseline. Model weight tended to fall off towards the end of the task performance period for HbO, and became negative for HbR about halfway through the task.

For better interpretability of the spatial weight vectors, we also estimated concentration differences in cortical space from model weights, here averaging out the weight time course (leaving only the spatial filters). Since spatial filters are not equivalent to activation maps, we use here the method of Haufe *et al* (2014) to first estimate a channel-space activation map (difference between experimental conditions) to which the model is optimally adapted, which we then localize using the same HD-DOT workflow as for block averages (figure 7).

The source activity inferred to be preferentially extracted by the classifier weights shows overall good agreement with regions expected to be activated in n-Back workload tasks (left and right DLPFC and VLPFC, cf. e.g. Owen *et al*, 2005, Cui *et al*, 2011, or Fishburn *et al* 2014), and show a focal region of largest absolute activation centered on two adjacent gyri. The inferred activation maps overlap with parallel results obtained from traditional block averaging based neuroimaging, seen in figure 9, but exhibit somewhat more negative weight for HbO.

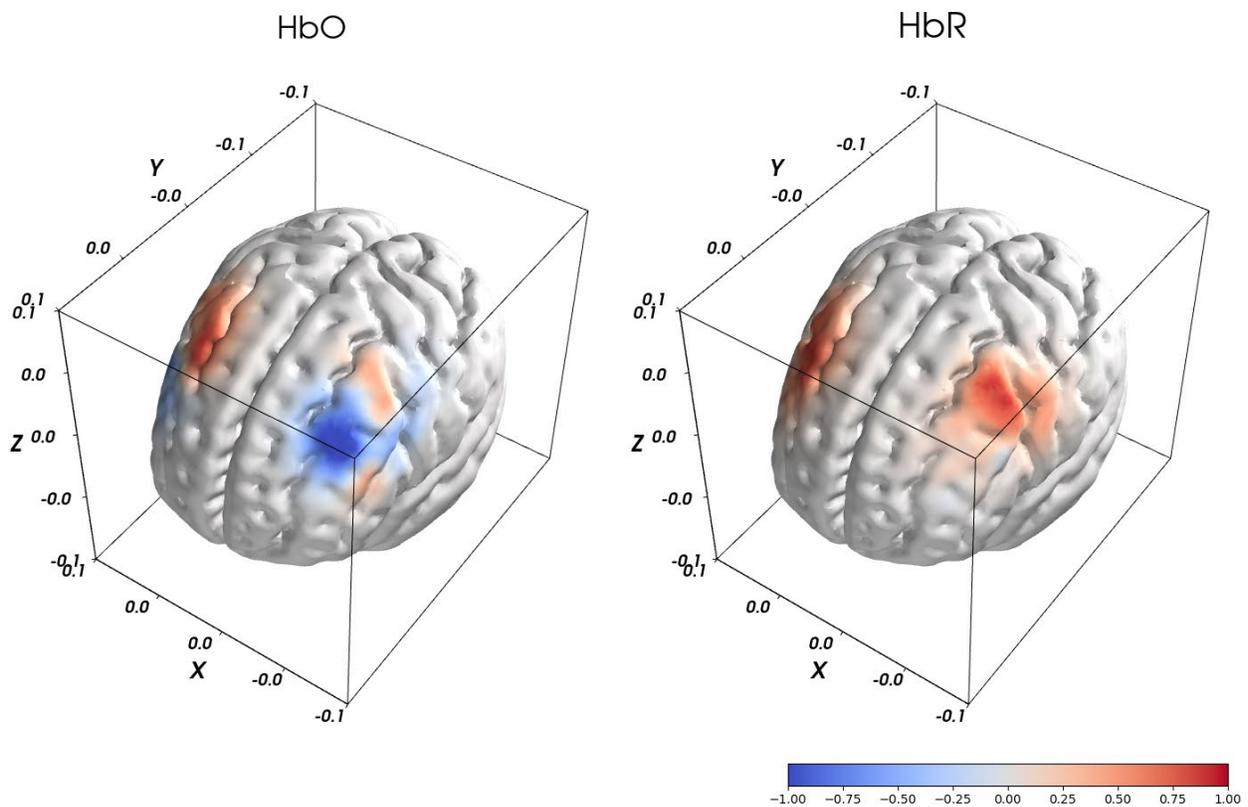

Figure 7. *Visualization of cortical maps of HbO (left) and HbR (right) concentration differences to which the model's linear spatial filter is maximally adapted. Model same as in* figure 5 *and* figure 6.

## Model Ablation Analysis

To assess the relevance of individual components of our model (e.g., regularization terms, preprocessing steps) we performed an ablation analysis, in which we start with the full model as analyzed in the previous section, and then successively remove model components, and re-measure the performance at each step. We utilized the setup with highest initial performance (OK session subset, n=0 vs 2 binary classification, 71.5 +/- 9.2%) as a starting point. The analysis (shown in figure 8) shows that each model aspect contributes to the overall performance, although some appear to contribute more than others. The largest differences were observed when disabling the recursive ZCA preprocessing step (6.4% absolute difference), followed by disabling cross-subject coupling via the MTL objective (4.6% absolute difference); the models trained without the cross-subject coupling terms are effectively trained only each individual participant's data. Note that these scores are somewhat order-dependent since a single regularization term often suffices to control a model's capacity and thus prevents overfitting, while additional terms mainly help refine or balance the prior assumptions imposed on the model (e.g., smoothness vs rank). Another large effect was attributable to bad-channel removal/interpolation (4.7% absolute difference). Also note some model features mostly impact variability across sessions rather than mean performance in this listing (e.g., the noise floor shift, bottom bar in figure 8).

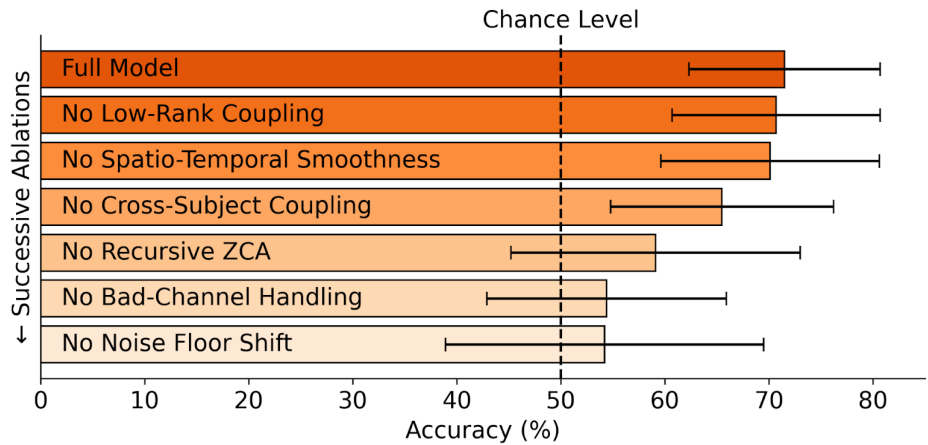

Figure 8. *Results from ablation analysis. Topmost bar shows classification accuracy of full model as proposed (tested on OK session subset, n=0 vs 2), with each bar below showing performance after disabling an additional model component, with removals being successive and cumulative (i.e., bottom-most processing pipeline is after all above model features have been disabled). Chance level is indicated by a dashed black line at 50% accuracy.*

## NIRS Block Averages

To aid interpretation of the machine learning results, specifically model weights and their tomographic maps and hemodynamic time courses, figure 9 shows grand-averaged contrasts of peak HbO/HbR activity averaged over the 5-25 s interval of the n-Back task data. We found there to be increased HbO activation for 2-back over 0-back in the left hemisphere VLPFC as well as in the right hemisphere middle and posterior DLPFC regions. HbR concentration shows an even more diffuse activation increase for 2-back compared to 0-back with increases in both left and right hemisphere middle and posterior DLPFC as well as right hemisphere VLPFC regions.

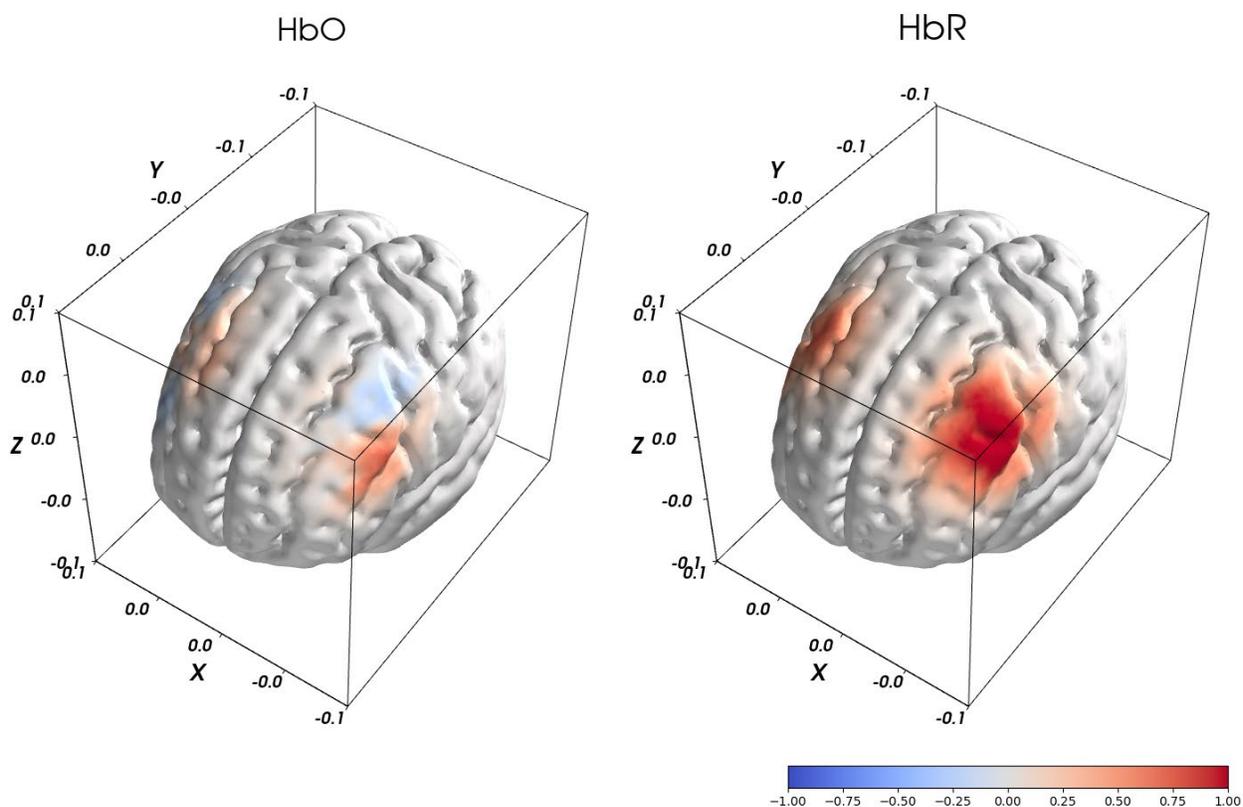

Figure 9. Cortical map of 2-back - 0-back contrast on the OK subset, averaged over the 5-25 s interval relative to task onset. Using the same HD-DOT imaging parameters as figure 7.

## Discussion

Our results represent one of the first machine learning analyses of NIRS data in the multi-thousand channel range, which poses considerable challenges due to the inherent high data dimensionality. We found that two existing NIRS methods proposed for higher-density NIRS data (Shin *et al*, 2017 and Ang *et al* 2014) do not readily scale to the data volume of the device tested here, which has several times the channel count as that analyzed in these prior studies (3198 dual-wavelength channels vs. 204 and 1024, respectively). This was evident in these methods not reaching their previously published performance (on other devices) nor the performance of the proposed method, which yielded significantly higher decoding accuracies. The proposed method had accuracies in the ca. 70% regime for the two-level n=0 vs 2 n-Back contrast, and around 60% between adjacent workload levels (n=0 vs 1 and n=1 vs 2). The existing methods yielded as much as 5% lower accuracy, depending on the chosen subject subset and subset of n-back levels to classify.

On the other hand, when comparing the proposed method to published results from comparable prior n-back workload studies (mostly on conventional low-density devices), we find the decoding performance to be in the same range as that of the current state-of-the-art methods for this problem. However, it is challenging to directly compare results across studies on even this simple task due to the many parameters affecting n-Back task difficulty, subject fatigue, skill and effort, and common pitfalls when comparing unlike evaluation procedures (e.g., randomized vs. block-wise within-session CV vs. leave-one-subject-out CV). In an attempt to filter the literature to the closest-matching experiment designs, and excluding studies where we could not rule out statistical double dipping (testing-on-the-training-data) hazards (e.g., overlapping train/test trials), we selected a few comparable results which we summarize below.

Herff *et al* (2014) report 55-60% accuracy depending on decoding window length (5-40s) when decoding adjacent workload levels (n=1 vs 2 or 2 vs 3) and 60-70% for the two-level difference of n=1 vs 3 using a linear method (LDA) in what appears to be a 10-fold blockwise cross-validation on 8-channel NIRS. These results resemble those seen in our data but are not quite at the same performance level.

Liu *et al*, (2017) report ca. 46.7% and 59.5% accuracy for 0 vs 1-back and 1 vs 2-back, respectively, and ca. 65.6% for 0 vs 2-back, using a standard setup consisting of an LDA applied to linear NIRS features from a 16ch headset, in a leave-one-out CV on fNIRS-only data, which is in line with the performance obtained with reproduced comparison methods in our analysis.

Huang *et al*, (2021) classify 0 vs 2 back both using subject-independent and subject-specific models in a custom-built 2-channel setup consisting of dual-slope frequency-domain NIRS probes, on a college cohort. The authors employ a rigorous evaluation scheme and, using a random forest approach on handcrafted features, report accuracies in the 65-77% range, depending on train/test demographics, training set size (up to 64 participants, larger datasets giving better performance), and on whether the method was subject-independent or subject-specific. Using $l_2$-regularized regression on the same features, they report accuracies of 60-67%.

Kesedžić *et al* (2021) classify 1 vs 2-back at 82% accuracy using an SCRDA classifier in their study, 2 vs 3-back at 64%, and 1-back vs. 3-back at 83% on 16-channel data using handcrafted features of sets of channels (e.g., slopes, averages, variances, ratios) and automated feature selection using a rigorous nested leave-one-subject out cross-validation. That study yielded the highest decoding performance we could find in a literature triage of clearly described and highly rigorous analyses of comparable n-back data. However, each n-Back trial is ca. twice as long as ours (75s vs 40s), likely yielding a better single-trial SNR when averaging over the trial duration, which we expect should boost decoding accuracy; this study also used a much lower-density montage than ours.

One challenge with high-density NIRS is the necessarily shorter duty cycle for each source optode, which results in lower SNR per channel at the same (safe) source brightness, or equivalently a spreading of available "light budget" from the sources over many more channels than in conventional low-density NIRS systems (e.g., 16+16 optodes with nearest-neighbor connectivity rather than all-to-all channels). As a result, highly reductive feature and channel selection is often not a useful analysis approach with such devices, and any successful method has to integrate over many channels to extract a high-SNR signal. All tested methods accomplish this by either heavy multi-channel averaging (Kesedžić *et al*, 2021) or by relying on the linear classifier to identify a weighted combination of channels (Shin *et al*, 2017; Ang *et al*, 2014), although both these prior methods nevertheless retain only a subset of channels (either a subset of source-detector separations, or a subset of most informative channel/time-window features).

In contrast, the proposed method follows the alternative philosophy of not selecting or reducing features beforehand, but instead leveraging all sources of information in the data, both across time and space, while reducing the effective degrees of freedom of the model using strong regularization (here, spatio-temporal smoothing and a low-rank assumption) in the classifier. Thus, while the information content is far more spread out across thousands of channels in high-density NIRS, our results suggest that it is possible for the ML model to aggregate this information effectively and match the performance of current state-of-the-art results obtained with lower-density devices. We also found our approach to be robust enough to tolerate some sessions with elevated noise levels (>15% CoV across channels) in both training and test data with minimal performance impact (ALL vs. OK subset).

A side effect of our choice is that the decoder stage does not require choosing multiple hand-tuned features such as time averages, slopes, and so forth, or carefully placed channel subsets based on the task and data at hand, but instead learns the necessary (linear) features and relevant channels from the data, while requiring the user to only select the overall epoch time window (here -15 to +45 s) and time resolution (here 0.2 Hz). As such, the method is task-invariant and should be readily applicable to other NIRS data without much modification. The effectiveness of this type of regularization is seen in the temporal model weights (figure 7), which are largely noise-free and resemble low-degree parametric curves. As a caveat, we observed that spatial smoothness was not as effective on this dataset as one may have expected, which may be attributable to the classifier learning "patchy" channel maps for this task as seen in figure 5 with somewhat complex inferred source activity distributions (figure 6).

Another strategy employed in the proposed method is to leverage as much data as is available to constrain the model, which here is other participants' data, although this could as well be multiple prior sessions from the same participant, or a combination of the two. The ablation analysis shows that the multi-task learning aspect contributed close to 5% absolute performance to the results (see figure 8). MTL or more generally transfer learning are relatively new frameworks (e.g., Zhang and Yang *et al*, 2021 for an overview on MTL) that have greatly advanced over the past decade in parallel to the deep learning revolution, and which offer a range of choices for leveraging auxiliary data effectively. Particularly when dealing with high-density NIRS, where high data dimensionality meets low trial counts, we believe these techniques are critical for leveraging all the information present in a given dataset and boosting available trial counts. Most of these techniques fall into the convex optimization framework (e.g., Boyd and Vandenberghe, 2004) and therefore come with strong global optimality guarantees, yet have so far only rarely been used in neural data analysis (e.g., Alamgir *et al* 2010, Wu *et al* 2020).

While outside the scope of this article, we also found ZCA (and its incremental variant for real-time processing) to be an effective general-purpose technique for higher-density NIRS preprocessing, and this was the single highest contributor to our model performance, as seen in the ablation analysis (see figure 8). Notably, a large performance gain due to ZCA is already seen without any advanced multi-subject learning or regularization, and despite there already being a conventional (CoV-based) artifact removal step in the pipeline. This may be attributable to ZCA both decorrelating (and thus separating) and shrinking activity in noisy sets of channels to unit variance, thereby limiting the impact of these channels on the classifier's performance, and doing so adaptively both at training and test time, especially in the presence of many channels. In exploratory analysis we had also seen improvements when ZCA was applied in the method of Shin *et al*, although this is not further explored here. A closely related method was also independently employed by Huang *et al* (2021).

## Conclusion

We present a new and effective machine learning strategy for analyzing high-channel NIRS data that is applicable not only to the n-Back task studied here, but which is also generic by design in that it requires no pre-selection of channels, time slices, or other task-specific features. The method demonstrates how fine-grained spatial and temporal information can be extracted effectively while leveraging multi-subject data, with or without per-subject adaptation. Our method has competitive performance compared with other tested methods on the same data, and when compared with published results for decoding cognitive workload in an n-Back task using other high or low density NIRS devices. Additionally, our method maintains good spatial and temporal interpretability due to the end-to-end linearity of all operations applied to optical density. Our results support that high-density NIRS devices such as the device tested here are viable tools for BCI applications and can achieve performance comparable to the state of the art when combined with an appropriate decoding method. An added benefit of high-density systems, combined with an appropriate modeling approach, is their affordance of high-resolution spatial model visualization and interpretation, as well as spatial imaging of brain activity using HD-DOT source localization or channel topographic maps, as done here. Such visualization may provide more fine-grained insights into potential biases of the model (e.g., leveraging unexpected brain regions or superficial artifacts) and can thus feature in safety or usability analysis for real-world deployment of such models.

## Acknowledgments

C K, S M, T M, and G H planned the study and designed the experiment, S M coded the experiment, C K and G H collected the data, C K performed the machine learning analysis, S M performed data quality analysis, and G H performed the block average and HD-DOT imaging analysis. All authors contributed to and edited the manuscript. We acknowledge Andrew Stewart for helpful input to the experiment design and contributions to the experiment implementation. This work was partially funded by Meta Reality Labs under research contract PO 70000160863. The authors declare no conflict of interest.

## References

Alamgir, M., Grosse–Wentrup, M. and Altun, Y., 2010, March. Multitask learning for brain-computer interfaces. In *Proceedings of the thirteenth international conference on artificial intelligence and statistics* (pp. 17-24). JMLR Workshop and Conference Proceedings.

Anaya, D., Batra, G., Bracewell, P., Catoen, R., Chakraborty, D., Chevillet, M., ... & Yin, A. (2023). Scalable, modular continuous wave functional near-infrared spectroscopy system (Spotlight). *Journal of Biomedical Optics*, 28(6), 065003-065003.

Ang, K.K., Yu, J. and Guan, C., 2014, August. Single-trial classification of NIRS data from prefrontal cortex during working memory tasks. In 2014 36th Annual International Conference of the IEEE Engineering in Medicine and Biology Society (pp. 2008-2011). IEEE.

Argyriou, A., Evgeniou, T. and Pontil, M., 2006. Multi-task feature learning. *Advances in Neural Information Processing Systems*, 19.

Ban, H.Y., Barrett, G.M., Borisevich, A., Chaturvedi, A., Dahle, J.L., Dehghani, H., DoValle, B., Dubois, J., Field, R.M., Gopalakrishnan, V. and Gundran, A., 2021, March. Kernel Flow: a high channel count scalable TD-fNIRS system. In *Integrated Sensors for Biological and Neural Sensing* (Vol. 11663, pp. 24-42). SPIE.

Beck, A. and Teboulle, M., 2009. A fast iterative shrinkage-thresholding algorithm for linear inverse problems. *SIAM Journal on Imaging Sciences*, 2(1), pp.183-202.

Bell, A. and Sejnowski, T.J., 1996. Edges are the 'Independent Components' of Natural Scenes. *Advances in Neural Information Processing Systems*, 9.

Boyd, S., Boyd, S.P. and Vandenberghe, L., 2004. *Convex optimization.* Cambridge university press.

Cui, X., Bray, S. and Reiss, A.L., 2010. Speeded near infrared spectroscopy (NIRS) response detection. *PLoS One*, 5(11), p.e15474.

Cui, X., Bray, S., Bryant, D.M., Glover, G.H. and Reiss, A.L., 2011. A quantitative comparison of NIRS and fMRI across multiple cognitive tasks. *Neuroimage*, 54(4), pp.2808-2821.

Dedovic, K., Renwick, R., Mahani, N.K., Engert, V., Lupien, S.J. and Pruessner, J.C., 2005. The Montreal Imaging Stress Task: using functional imaging to investigate the effects of perceiving and processing psychosocial stress in the human brain. *Journal of Psychiatry and Neuroscience*, 30(5), pp.319-325.

Delpy, D.T., Cope, M., van der Zee, P., Arridge, S., Wray, S. and Wyatt, J.S., 1988. Estimation of optical pathlength through tissue from direct time of flight measurement. *Physics in Medicine & Biology*, 33(12), p.1433.

Dempster, A.P., Laird, N.M. and Rubin, D.B., 1977. Maximum likelihood from incomplete data via the EM algorithm. *Journal of the Royal Statistical Society: Series B (Methodological)*, 39(1), pp.1-22.

Enders, C.K., 2022. *Applied missing data analysis.* Guilford Publications.

Fishburn, F.A., Ludlum, R.S., Vaidya, C.J. and Medvedev, A.V., 2019. Temporal derivative distribution repair (TDDR): a motion correction method for fNIRS. *NeuroImage*, 184, pp.171-179.

Fishburn, F.A., Norr, M.E., Medvedev, A.V. and Vaidya, C.J., 2014. Sensitivity of fNIRS to cognitive state and load. *Frontiers in Human Neuroscience*, 8, p.76.

Fonov, V.S., Evans, A.C., McKinstry, R.C., Almli, C.R. and Collins, D.L., 2009. Unbiased nonlinear average age-appropriate brain templates from birth to adulthood. *NeuroImage*, (47), p.S102.

Guo, Y., Hastie, T. and Tibshirani, R., 2007. Regularized linear discriminant analysis and its application in microarrays. *Biostatistics*, 8(1), pp.86-100.

Haufe, S., Meinecke, F., Görgen, K., Dähne, S., Haynes, J.D., Blankertz, B. and Bießmann, F., 2014. On the interpretation of weight vectors of linear models in multivariate neuroimaging. *Neuroimage*, 87, pp.96-110.

Herff, C., Heger, D., Fortmann, O., Hennrich, J., Putze, F. and Schultz, T., 2014. Mental workload during n-back task—quantified in the prefrontal cortex using fNIRS. *Frontiers in Human Neuroscience*, 7, p.935.

Huang, Z., Wang, L., Blaney, G., Slaughter, C., McKeon, D., Zhou, Z., Jacob, R. and Hughes, M.C., 2021. The Tufts fNIRS Mental Workload Dataset & Benchmark for Brain-Computer

Interfaces that Generalize. Available at: https://openreview.net/forum?id=QzNHE7QHhut (Accessed: Nov 16 2022)

Huber, P.J., 1996. *Robust statistical procedures.* Society for Industrial and Applied Mathematics.

Kesedžić, I., Šarlija, M., Božek, J., Popović, S. and Ćosić, K., 2020. Classification of cognitive load based on neurophysiological features from functional near-infrared spectroscopy and electrocardiography signals on n-back task. IEEE Sensors Journal, 21(13), pp.14131-14140.

Kirchner, W.K., 1958. Age differences in short-term retention of rapidly changing information. *Journal of experimental psychology*, 55(4), p.352.

Kothe, C., Boulay, C., Stenner, T., Medine, D., and Delorme, A., 2012 The Lab Streaming Layer. Available at: https://github.com/labstreaminglayer (Accessed: Nov 15 2022)

Kriegeskorte, N., Simmons, W. K., Bellgowan, P. S., & Baker, C. I. (2009). Circular analysis in systems neuroscience: the dangers of double dipping. *Nature Neuroscience*, 12(5), 535-540

Lavie, N., 1995. Perceptual load as a necessary condition for selective attention. Journal of Experimental Psychology: *Human Perception and Performance*, 21(3), p.451.

Lavie, N., 2010. Attention, distraction, and cognitive control under load. *Current Directions in Psychological Science*, 19(3), pp.143-148.

Madsen, S.J. ed., 2012. *Optical methods and instrumentation in brain imaging and therapy (Vol. 3).* Springer Science & Business Media.

McDonald, A.M., Pontil, M. and Stamos, D., 2016. New perspectives on k-support and cluster norms. *The Journal of Machine Learning Research*, 17(1), pp.5376-5413.

Nesterov, Y., 2013. Gradient methods for minimizing composite functions. *Mathematical Programming*, 140(1), pp.125-161.

Niendam, T.A., Laird, A.R., Ray, K.L., Dean, Y.M., Glahn, D.C. and Carter, C.S., 2012. Meta-analytic evidence for a superordinate cognitive control network subserving diverse executive functions. *Cognitive, Affective, & Behavioral Neuroscience*, 12(2), pp.241-268.

Okuta, R., Unno, Y., Nishino, D., Hido, S. and Loomis, C., 2017. CuPy: A Numpy-Compatible Library for NVIDIA GPU Calculations. 31st Conference on Neural Information Processing Systems, 151(7).

Oquab, M., Rapin, J., Teytaud, O. and Cazenave, T., 2019. Parallel Noisy Optimization in Front of Simulators: Optimism, Pessimism, Repetitions, Population Control. In *Workshop Data-driven Optimization and Applications at CEC*.

Owen, A.M., McMillan, K.M., Laird, A.R. and Bullmore, E., 2005. N-back working memory paradigm: A meta-analysis of normative functional neuroimaging studies. *Human Brain Mapping*, 25(1), pp.46-59.

Pal, K.K. and Sudeep, K.S., 2016, May. Preprocessing for image classification by convolutional neural networks. In 2016 IEEE International Conference on Recent Trends in Electronics, Information & Communication Technology (RTEICT) (pp. 1778-1781). IEEE.

Peck, R. and Van Ness, J., 1982. The use of shrinkage estimators in linear discriminant analysis. *IEEE Transactions on Pattern Analysis and Machine Intelligence*, (5), pp.530-537.

Schweiger, M. and Arridge, S.R., 2014. The Toast++ software suite for forward and inverse modeling in optical tomography. *Journal of Biomedical Optics*, *19*(4), p.040801.

Shin, J., Kwon, J., Choi, J. and Im, C.H., 2017. Performance enhancement of a brain-computer interface using high-density multi-distance NIRS. Scientific Reports, 7(1), pp.1-10.

Tikhonov, A.N., Goncharsky, A.V., Stepanov, V.V. and Yagola, A.G., 1995. *Numerical methods for the solution of ill-posed problems (Vol. 328).* Springer Science & Business Media.

Tremblay, J., Martínez-Montes, E., Vannasing, P., Nguyen, D.K., Sawan, M., Lepore, F. and Gallagher, A., 2018. Comparison of source localization techniques in diffuse optical tomography for fNIRS application using a realistic head model. *Biomedical optics express*, 9(7), pp.2994-3016.

Varoquaux, G., Raamana, P.R., Engemann, D.A., Hoyos-Idrobo, A., Schwartz, Y. and Thirion, B., 2017. Assessing and tuning brain decoders: cross-validation, caveats, and guidelines. *NeuroImage*, 145, pp.166-179.

Vidal-Rosas, E.E., Zhao, H., Nixon-Hill, R.W., Smith, G., Dunne, L., Powell, S., Cooper, R.J. and Everdell, N.L., 2021. Evaluating a new generation of wearable high-density diffuse optical tomography technology via retinotopic mapping of the adult visual cortex. *Neurophotonics*, 8(2), p.025002.

Wheelock, M.D., Culver, J.P. and Eggebrecht, A.T., 2019. High-density diffuse optical tomography for imaging human brain function. *Review of Scientific Instruments*, 90(5), p.051101.

Wu, D., Xu, Y. and Lu, B.L., 2020. Transfer learning for EEG-based brain–computer interfaces: A review of progress made since 2016. *IEEE Transactions on Cognitive and Developmental Systems*, 14(1), pp.4-19.

Zander, T.O. and Kothe, C., 2011. Towards passive brain–computer interfaces: applying brain–computer interface technology to human–machine systems in general. *Journal of Neural Engineering*, 8(2), p.025005.

Zhang, Y. and Yang, Q., 2021. A survey on multi-task learning. *IEEE Transactions on Knowledge and Data Engineering*.

Zhao, H., Frijia, E.M., Rosas, E.V., Collins-Jones, L., Smith, G., Nixon-Hill, R., Powell, S., Everdell, N.L. and Cooper, R.J., 2021. Design and validation of a mechanically flexible and ultra-lightweight high-density diffuse optical tomography system for functional neuroimaging of newborns. *Neurophotonics*, 8(1), p.015011.